\documentclass[reprint,amsmath,amssymb,aps,prl,superscriptaddress,nofootinbib]{revtex4-2}

\usepackage{graphicx,color} 
\usepackage{dcolumn}
\usepackage{bm}
\usepackage{mathrsfs}
\usepackage{physics}
\usepackage{hyperref}
\usepackage{xcolor}
\usepackage[whole]{bxcjkjatype} 
\usepackage{ulem}
\usepackage{txfonts}

\newcommand{\ds}{\displaystyle}

\begin{document}

\preprint{QME}

\title{Unexpected linear conductivity in Landau-Zener model: limitations and improvements  of the relaxation time approximation in the quantum master equation}

\author{Ibuki Terada}
\affiliation{
Department of Physics, Ritsumeikan University, Shiga 525-8577, Japan
}
\author{Sota Kitamura}
\affiliation{
Department of Applied Physics, The University of Tokyo, Hongo, Tokyo, 113-8656, Japan
}
\author{Hiroshi Watanabe}
\affiliation{
Research Organization of Science and Technology, Ritsumeikan University, Shiga 525-8577, Japan
}
\author{Hiroaki Ikeda}
\affiliation{
Department of Physics, Ritsumeikan University, Shiga 525-8577, Japan
}

\date{\today}

\begin{abstract}
The nonequilibrium steady states of quantum materials have many challenges. 
Here, we highlight issues with the relaxation time approximation (RTA) for the DC conductivity in insulating systems. The RTA to the quantum master equation (QME) is frequently employed as a simple method, yet this phenomenological approach is exposed as a fatal approximation, displaying metallic DC conductivity in insulating systems within the linear response regime.
We find that the unexpected metallic behavior is caused by the fact that the density matrix in the RTA incompletely incorporates the first order of the external field.
To solve this problem, we have derived a new calculation scheme based on the QME that ensure correct behavior in low electric fields.
Our method reproduces well the overall features of the exact electric currents in the whole field region. It is not time-consuming, and its application to lattice systems is straightforward. This method will encourage progress in this research area as a simple way to more accurately describe nonequilibrium steady states.
\end{abstract}

\maketitle

{\it Introduction}---
Field-induced phenomena in quantum systems under strong electric fields have attracted much attention with the recent developments in laser technology~\cite{Aoki,Basov,Oka,Eckardt,Torre,Morimoto}. 
The electromagnetic responses of condensed-matter systems beyond the linear response regime lead to a variety of intriguing phenomena, including bulk photovoltaic effects ~\cite{Boyd,Bloembergen,Sturman,Nie,Shi,deQuilettes} and nonreciprocal transport of quantum geometric origin~\cite{Kitamura,Kitamura2,Takayoshi,Suzuki,Sodemann}. Laser light with stronger intensities triggers photo-induced phase transitions as a result of nonperturbative quantum effects, which is experimentally demonstrated in the ultrafast time scale and investigated theoretically~\cite{Nasu,Basov2,Oka2,Kitagawa,Mclver,Kirilyuk,Takayoshi2,Takayoshi3,MSato}.

When the driving field becomes strong, dissipation to environment plays a vital role in determining the distribution function far from equilibrium, which is generally difficult and has a long history of research~\cite{Weiss,Barreiro,Vandecasteele,Fang,Li,Okamoto,Sugimoto,Heidrich-Meisner,Ikeda}. 
Theoretical description of open dissipative systems necessitates the nonequilibrium Green's function ~\cite{Aoki,Blandin,Jauho,Onoda,Han} or the density matrix ~\cite{Sipe,Sipe2,Ventura,Passos} as a fundamental quantity, rather than the wave function. 
The nonequilibrium Green's function can be calculated using the diagrammatic approach formulated on the Schwinger-Keldysh contour, while the density matrix is calculated on the basis of the quantum master equation (QME) ~\cite{Breuer,Alicki}. 
Recently, nonperturbative modulation of the distribution function due to the Landau-Zener tunneling ~\cite{Zener,Davis,Dykhne} is formulated by combining the nonequilibrium Green's functions approach with analytic methods~\cite{Kitamura2}. There, it was discussed that systems with broken inversion symmetry lead to interesting nonperturbative phenomena such as tunnel spin current and nonreciprocal current.
Such Green's function method is a powerful tool but requires the treatment of cumbersome dynamical phases in applications to lattice systems. The development of QME methods as a complementary method is therefore considered important. 

The relaxation time approximation (RTA) is the simplest approximation to describe the nonequilibrium steady state in the QME approach~\cite{Sato,Nuske,Sato2}. 
It has been frequently used in the context of the semi-classical Boltzmann equation and the semiconductor Bloch equation. 
Quite recently, however, it has been pointed out that the RTA has a problem in its application to nonlinear optics~\cite{Passos,Michishita}. 
We show here that, in addition to this problem, the RTA treatment in insulating systems involves a fatal failure that the DC conductivity exhibits metallic behavior in the linear response regime, despite the absence of Fermi surfaces. 
The reason why this has been overlooked until now is that 
transport properties are mainly studied in the perturbative regime (i.e., the DC conductivity of metals or the AC conductivity of insulators), whereas 
the tunneling current is a purely nonperturbative response seen only under strong fields.

In this letter, we demonstrate the failure of the RTA using the Landau-Zener model as a minimal example. 
We find that the fatal metallic behavior originates from the fact that the density matrix in the RTA incorporates the first order of the external field $E$ in an incomplete form. 
On the basis of the QME, we derive a calculation scheme to ensure correct behavior in low electric fields by sequentially incorporating the perturbation correction of the external field. 
Our method reproduces well the overall features of electric currents in the whole field region.
It is not time-consuming, and its computational time is comparable to RTA calculations. 

{\it Relaxation Time Approximation}---
To demonstrate a fatal failure of the RTA, let us consider a two-band system with a finite gap for simplicity.
With two eigenenergies $\varepsilon_{k\pm}$, the band gap at each $k$ point is given by $\Delta_k=\varepsilon_{k+}-\varepsilon_{k-}>0$.
We introduce the DC electric field $E$ via the Peierls substitution, $k\to k-eEt$ (hereafter $e=1$).
Following Ref.\,\cite{Kitamura2}, we introduce the snapshot basis $|\Phi_{k\alpha}(t)\rangle=|u_{\alpha,k-Et}\rangle e^{-i\Theta_{\alpha}(t)}$ ($\alpha=\pm$) with $|u_{\alpha,k}\rangle$ being the eigenstate of the snapshot Hamiltonian $H(k)$. 
The phase factor $\Theta_\alpha(t)$ consists of the dynamical and Berry phases~\cite{comment}.
In this basis, the QME for the density matrix $[\rho_k(t)]_{\alpha\beta}$ is written in the $2\times2$ matrix form as
\begin{align}\label{eq:QME_origin}
	\frac{d\rho_k(t)}{dt}=-i\big[ \mathcal{W}_k(t),\rho_k(t)\big]+\mathcal{D}\big[\rho_k(t)\big],
\end{align}
where $[\mathcal{W}_k(t)]_{\alpha\alpha}=0$, $[\mathcal{W}_k(t)]_{+-}=[\mathcal{W}_k(t)]^*_{-+}\equiv W_k(t)=E\langle u_{+,k-Et}|i\partial_k|u_{-,k-Et}\rangle e^{i\Theta_+(t)-i\Theta_-(t)}$  denotes the transition dipole matrix elements.
Here the dissipation to environment is described by $\mathcal{D}\big[\rho_k(t)\big]$,
whose form in the RTA is given by 
\begin{align}\label{eq:coll_pheno}
	\big[\mathcal{D}_k[\rho_k(t)]\big]_{\alpha\beta}=\frac{f_D(\varepsilon_{k\alpha}(t))\delta_{\alpha\beta}-\big[\rho_k(t)\big]_{\alpha\beta}}{\tau_{\alpha\beta}},
\end{align}
where $f_D$ is the Fermi-Dirac distribution function, and $\varepsilon_{k\alpha}(t)=\varepsilon_{k-Et,\alpha}$. 
$\tau_{++}=\tau_{--}=\tau_1$ and $\tau_{+-}=\tau_{-+}=\tau_2$ denote the longitudinal and transverse relaxation time, respectively.

By solving Eq.~(\ref{eq:QME_origin}), we obtain the density matrix of the nonequilibrium steady state.
Then the electric current $J(t)$ is calculated as $J(t)=-e\int\frac{dk}{2\pi}\text{Tr}[\partial_k H(k) \rho_k(t)]$, which can be decomposed into intra- and interband contributions as  
\begin{align}\label{J_contri}
	&J(t)=J_{\rm{intra}}(t)+J_{\rm{inter}}(t),\\
	&J_{\rm{intra}}(t)
	=-\sum_{\alpha=\pm}\int\frac{dk}{2\pi}\frac{\partial\varepsilon_{k\alpha}(t)}{\partial k}\big[\rho_k(t)\big]_{\alpha\alpha},\\
	&J_{\rm{inter}}(t)
	=2{\rm Im}\left[\int\frac{dk}{2\pi}\frac{W_k(t)\Delta_k(t)}{ E}\big[\rho_k(t)\big]_{-+}\right].
	\label{eq:J_inter}
\end{align}
The electric current in the nonequilibrium steady states is obtained as the long-time limit of $J(t)$.

Let us exemplify the electric current in the RTA by a numerical calculation for the Landau-Zener model
\begin{equation}\label{LZ_C}
	H(k)=\left(\begin{matrix}vk&\delta\\ \delta&-vk\end{matrix}\right),
\end{equation}
where $2\delta$ corresponds to the band gap. 
In the absence of the dissipation, the tunneling probability in the $t\to\infty$ limit is exactly given by $P_{\rm{LZ}}=\exp({-\pi E_{\rm th}/E})$ with $E_{\rm th}=\delta^2/v$ [the dashed line of the main of Fig.~\ref{fig:RTA}(a)]~\cite{Zener,Grandi}.
Figure~\ref{fig:RTA}(a) shows the field dependence of the RTA electric current $J^{\rm RTA}$ with $\tau_1=\tau_2=\tau$ at the temperature $T=0.01\delta$. 
One can see a remarkable increase of the current $J^{\rm RTA}$ at around $E=0.5E_{\rm th}$, consistent with the generation of tunneling carriers. 
Surprisingly, an unexpected linear $E$ dependence with the slope increasing with $\tau^{-1}$ is observed in the low $E$ regime, indicating metallic behavior even though the system is actually an insulator. 
What is responsible for this finite linear dependence? 
To clarify the origin, in Fig.~\ref{fig:RTA}(b), we separately show the contributions from $J^{\rm RTA}_{\rm intra}$ (dashed line) and $J^{\rm RTA}_{\rm inter}$ (dotted line) for $\tau^{-1}=0.02\delta$. One can see that the $E$ linear behavior comes from the interband contribution $J^{\rm RTA}_{\rm inter}$. 

\begin{figure}[t]
\centering
\includegraphics[width=7.5cm]{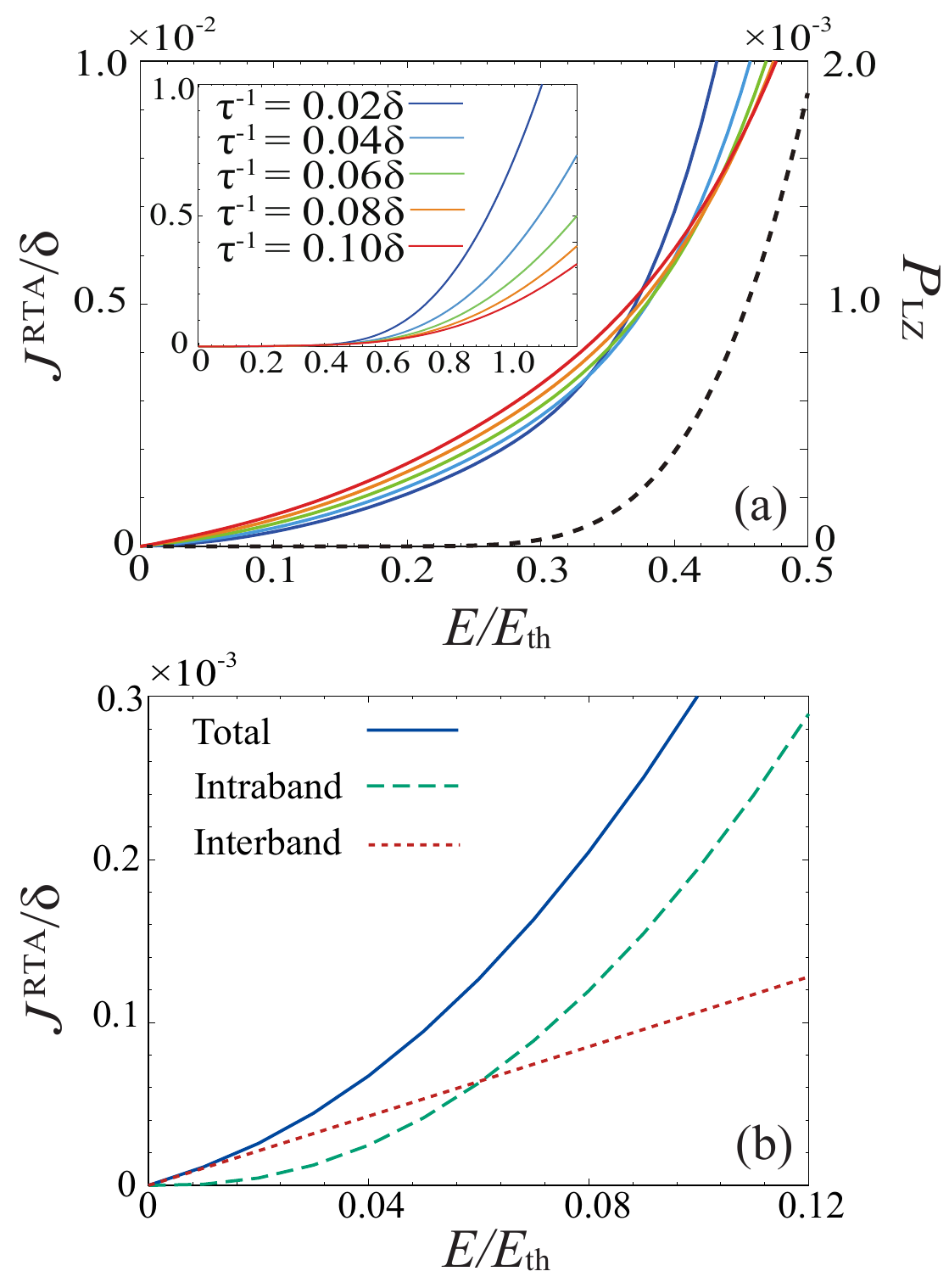}
\caption
{(a) Field dependence of the DC current $J^{\rm RTA}$ in the RTA at $T=0.01\delta$. The inset depicts $J^{\rm RTA}$ in a wide range. Solid lines denote the DPA currents for several $\tau^{-1}$. 
The dotted line represents the tunneling probability $P_{\rm{LZ}}$. One can see the exponential behavior in the inset, but encounter an unexpected linear behavior at a low $E$ limit in the main panel.
(b) Intraband $J^{\rm RTA}_{\rm intra}$ (dashed line) and interband $J^{\rm RTA}_{\rm inter}$ (dotted line) contributions of $J^{\rm RTA}$ at $\tau^{-1}=0.02\delta$ as a function of $E/E_{\rm th}$. $J^{\rm RTA}_{\rm inter}$ shows unphysical linear $E$ behavior. 
}\label{fig:RTA}
\end{figure}

In fact, the density matrix in the low $E$ limit can be calculated as 
\begin{align}
	\label{pheno_dia}
	&\left[\rho_k(t)\right]_{\alpha\alpha}\sim
	f_D(\varepsilon_{k\alpha}(t))+E\tau_1\frac{\partial f_D(\varepsilon_{k\alpha}(t))}{\partial k},\\
	\label{eq:pheno_offdia}
	&\left[\rho_k(t)\right]_{+-}\sim -\frac{W_k(t)}{\Delta_k(t)-i\tau_2^{-1}}\delta f_k(t)
\end{align}
with $\delta f_k(t)=f_D(\varepsilon_{k-}(t))-f_D(\varepsilon_{k+}(t))$, by solving Eq.~\eqref{eq:QME_origin} in a perturbative manner.
Then, within the linear response, the electric conductivities $\ds \sigma^{\rm RTA}=\lim_{E\to0} J^{\rm RTA}/E$ are given by
\begin{align}\label{Drude}
	\sigma^{\rm RTA}_{\rm intra}& =\tau_1\sum_{\alpha=\pm}\int\frac{dk}{2\pi}
	\left(\frac{\partial\varepsilon_{k\alpha}}{\partial k}\right)^2
	\left(-\frac{\partial f_D}{\partial \varepsilon_{k\alpha}}\right), \\
	\label{sig_Pol}
	\sigma^{\rm RTA}_{\rm inter}& 
	=2\tau_2^{-1}\int\frac{dk}{2\pi}\frac{|\langle u_{+,k}|i\partial_k|u_{-,k}\rangle|^2\Delta_k}{\tau_2^{-2}+\Delta_k^2}\delta f_k.
\end{align}
In the insulating case, the intraband conductivity $\sigma^{\rm RTA}_{\rm intra}$ vanishes at $T=0$, which is consistent with the Boltzmann theory.
On the other hand, the interband conductivity $\sigma^{\rm RTA}_{\rm inter}$ does not vanish even in the insulating case, which gives the slope of $J^{\rm RTA}_{\rm inter}$ at $E\to0$. 
This term increases roughly in proportion to damping $\tau_2^{-1}$. 
The small but finite linear conductivity, which exists despite the absence of Fermi surfaces, is a failure of the RTA in the QME. 
Although the RTA is useful to describe the nonequilibrium steady state in a simple manner, it has been recently argued that this approximation is also problematic for the optical response~\cite{Passos,Michishita}, which is another problem different from the unphysical behavior observed here for the DC current.

As we will show later, this problem comes from the presence of $i\tau_2^{-1}$ in the denominator of Eq.~(\ref{eq:pheno_offdia}). 
This is due to the fact that the RTA does not properly incorporate the first-order contribution of the electric field $E$.  
In our formalism beyond the RTA, this term in the denominator vanishes [cf. Eq.~(\ref{eq:pheno_offdia_corrected})]. 
This $i\tau_2^{-1}$ term also affects the Hall current, with which the Hall conductivity does not quantize within the RTA \cite{Sato}.

{\it Beyond RTA}---
To resolve the failure of the RTA, here we consider to couple the two-band insulator to a fermionic reservoir ~\cite{Aoki,Buttiker} within the QME formalism, and derive the dissipation term $\mathcal{D}[\rho_k (t)]$ microscopically.
We start with the Born-Markov master equation~\cite{Breuer,DelRe}
\begin{equation}\label{eq:QME}
	\frac{d \tilde{\rho}_{k}(t)}{dt}=
	-\int_{-\infty}^t\mathrm{Tr_B}\left[ \tilde{H}_{I,k}(t),\left[\tilde{H}_{I,k}(s),\tilde{\rho}_k(t)\otimes \tilde{\rho}_B\right]\right]ds,
\end{equation}
where $\tilde{\rho}_{k}(t)$ is the reduced density operator and $\tilde \rho_B$ is the thermal density operator of the bath, respectively.
Here the tilde on operators denote the interaction picture. 
$\tilde{H}_{I,k}(t)$ represents the interaction term between the system and the bath, $\tilde H_{I,k}(t)=\sum_{\sigma p}V_p\tilde b^{\dag}_{k\sigma p}(t)\tilde c_{k\sigma}(t)+\text{H.c.}$,
where $\tilde c_{k\sigma}(t)$ and $\tilde b_{k\sigma p}(t)=\tilde b_{k\sigma p}(0)e^{-i\omega_p t}$ are respectively the annihilation operator of an electron in the insulator and reservoir, with momentum $k$ and pseudospin $\sigma$. 
$\mathrm{Tr_B}\left[\cdots\right]$ means tracing out the bath degrees of freedom. 
We impose the broadband condition for the spectral density of the fermionic reservoir as
\begin{align}\label{eq:BBC}
	\sum_p \pi |V_p|^2\,\delta(\omega-\omega_p)=\Gamma~~({\rm const.}).
\end{align}

The key point of our formalism is to express Eq.~(\ref{eq:QME}) in the snapshot basis by introducing the transformed field operator
\begin{equation}
\tilde{\psi}_{k\alpha}(t)
=\sum_\sigma\braket{\Phi_{k\alpha}(t)}{\sigma}\tilde{c}_{k\sigma}(t),
\end{equation}
and evaluate the integral in Eq.~\eqref{eq:QME} with employing the adiabatic perturbation theory~\cite{Kitamura,Kitamura2}.
Specifically, in the adiabatic limit $E\to0$, we can incorporate changes in the dynamical phase as 
\begin{align}\label{app_W}
    W_{k}(t-s)&\sim e^{-i\Delta_{k}(t)s}W_{k}(t),\\
    |\Phi_{k\alpha}(t-s)\rangle&\sim e^{i\varepsilon_{k\alpha}(t)s}|\Phi_{k\alpha}(t)\rangle.
\end{align}
We further include the change of the parameter $\Delta k=-Es$ in a perturbative manner, 
such that the resultant QME is exact up to $O(E)$ while maintaining the unitarity of the time evolution operator up to $O(E^2)$. We call this approximation the dynamical phase approximation (DPA).

After all, we obtain the dissipation term in the DPA as ${\cal D}[\rho_k(t)]={\cal D}_0+{\cal D}_1+{\cal D}_2$, where
\begin{align}
	&\big[{\cal  D}_0\big]_{\alpha\beta}
	=-2\Gamma\big([\rho_k(t)]_{\alpha\beta}-f_D(\varepsilon_{k\alpha}(t))\delta_{\alpha\beta}\big),\\
	&\big[{\cal  D}_1\big]_{\alpha\beta}
	=-2\Gamma\frac{[{\cal W}_k(t)]_{\alpha\beta}}{\Delta_k(t)}\delta f_k(t),\label{eq:RME-1st}\\
	&\big[{\cal  D}_2\big]_{\alpha\beta}=2\Gamma\alpha\frac{|W_k(t)|^2}{\Delta_k^2(t)}
	\bigg(\delta f_k(t)+\Delta_k(t)f_D^\prime(\varepsilon_{k,-\alpha}(t))\bigg)\delta_{\alpha\beta}.
\end{align}
The subscript $n$ of ${\cal D}_n$ denotes the perturbation order of $W_k$~\cite{suppl}. Note that the expression here is simplified one under the particle-hole symmetry,  $\varepsilon_{k+}(t)=-\varepsilon_{k-}(t)$. 
Here the zeroth order term ${\cal D}_0$ corresponds to the RTA with $\tau_1=\tau_2=1/2\Gamma$,
while ${\cal D}_1, {\cal D}_2$ describe the field-induced correction terms.
Specifically, in the presence of Eq.~\eqref{eq:RME-1st}, the low-$E$ expression for $[\rho_{k}(t)]_{+-}$ is replaced from Eq.~\eqref{eq:pheno_offdia} into
\begin{equation}
[\rho_{k}(t)]_{+-}\sim-\dfrac{W_{k}(t)}{\Delta_{k}(t)}\delta f_{k}(t),\label{eq:pheno_offdia_corrected}
\end{equation}
with which unphysical linear term in $J^{\rm RTA}_{\rm inter}$ [Eq.~\eqref{eq:J_inter}] is completely canceled out.

\begin{figure}[t]
\centering
\includegraphics[width=7.5cm]{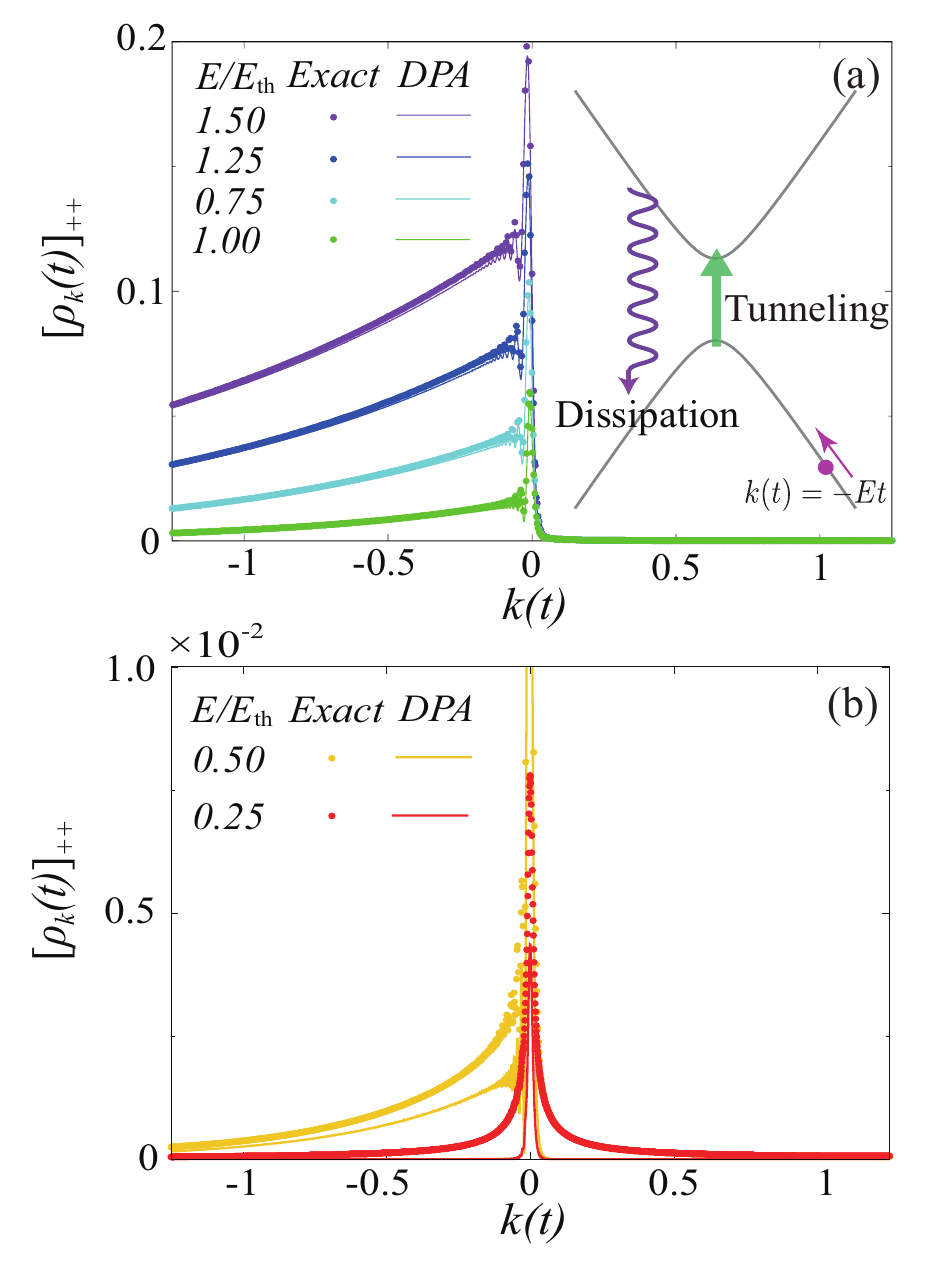}
\caption{Occupation number of a single electron in the upper band. Solid lines and filled circles denote DPA results and the numerically exact results~\cite{Kitamura2}, respectively. The schematic figure represents the electron dynamics in the Landau-Zener model. The DPA results are almost consistent with the exact results. The small deviation in due to the damping-induced excitation inherent in open systems.}\label{fig:2-2}
\end{figure}

Now, let us compare the DPA calculation with the numerically exact calculation~\cite{Kitamura2}. Figures~\ref{fig:2-2}(a) and (b) show the occupation numbers of the upper band at $\tau^{-1}=2\Gamma=0.02\delta$ and $T=0.01\delta$ as a function of $k(t)=-Et$. 
Electrons starting from $t=t_0<0~(k>0)$ are excited by Landau-Zener tunneling as they pass through the gap minimum at $t=0~(k=0)$. Then, at $t>0~(k<0)$, the excited electrons decay with a relaxation time $\tau$. 
The DPA results are consistent with the exact result at higher temperatures and stronger electric fields.
The small deviation in the low-$E$ regime is due to the damping-induced excitation inherent to the present fermionic reservoir, which is only partially included in the DPA.
While this is incorrect from the viewpoint of the rigorous treatment of the fermionic reservoir, it could be considered as an advantage since this particular excitation should be absent for ideal environment.

\begin{figure}[t]
\centering
\includegraphics[width=7cm]{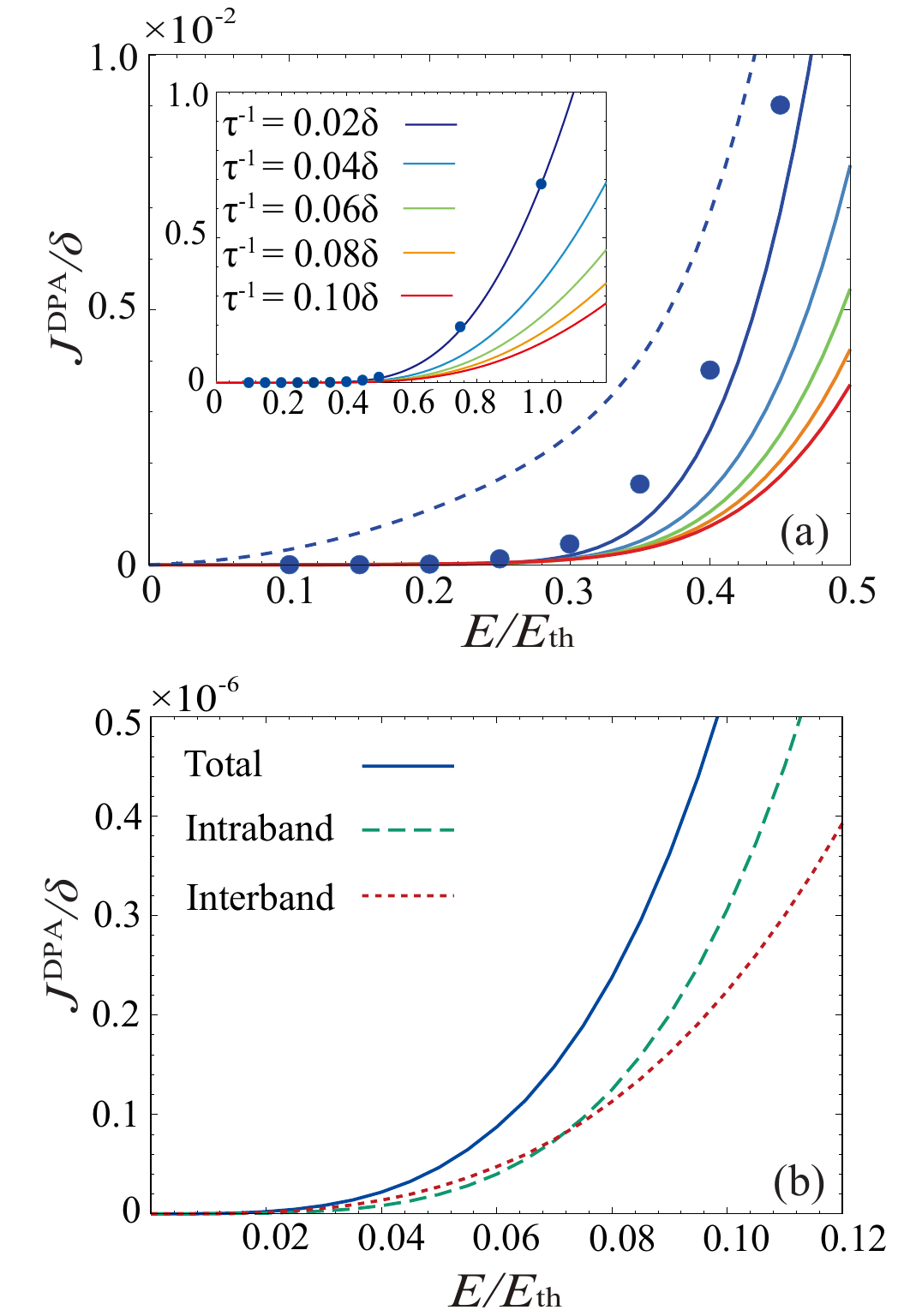}
\caption{(a) Field dependence of the DC current $J^{\rm DPA}$ in the DPA. The inset depicts $J^{\rm RTA}$ in the wide range. 
Solid lines denote the DPA currents for several $\tau^{-1}$, the dashed line represents the RTA current at $\tau^{-1}=0.02\delta$ and the filled circles the result of the numerically exact result. In the inset, one can see that the DPA result reproduces the exact result well. $J^{\rm DPA}$ almost vanishes at $E<0.2E_{\rm th}$. We can verify in (b) that $J^{\rm DPA}$ has no linear $E$ terms with an accuracy of less than $10^{-7}\delta$ at $E < 0.06E_{\rm th}$.
At intermediate region $E\sim 0.4E_{\rm th}$, $J^{\rm DPA}$ deviates from the exact result by $\sim 10^{-3}\delta$ but roughly increases with the tunneling probability $P$, implying that the excited carriers are carrying the electric current.
}\label{fig:3-2}
\end{figure}


Finally, we illustrate the field dependence of the electric current in the DPA in Fig.~\ref{fig:3-2}. 
Solid lines in Fig.~\ref{fig:3-2}(a) denote the DPA currents for several damping parameters, the dotted line represents the RTA current at $\tau^{-1}=0.02\delta$, and the filled circles are the numerically exact results. It is clear that the RTA fails to capture qualitative features of the exact result. On the other hand, as shown in the inset of Fig.~\ref{fig:3-2}(a), the DPA result well describes the exact one. 
The most significant improvement is the disappearance of the linear $E$ dependence in the interband contribution. This is due to the fact that the $i\tau^{-1}$ term in the off-diagonal term of the density matrix in the RTA is completely canceled by correctly treating the electric field up to the first order. This also improves the behavior of the intraband contribution, where the exponential behavior is more pronounced. As a result, the total current in the low-field region is also greatly improved and shows exponential behavior. Indeed, $J^{\rm DPA}$ almost vanishes at $E<0.2E_{\rm th}$. It can be verified in Fig.~\ref{fig:3-2}(b) that $J^{\rm DPA}$ has no linear $E$ terms with an accuracy of less than $10^{-7}\delta$ at $E < 0.06E_{\rm th}$. 
At intermediate region $E\sim 0.4E_{\rm th}$, $J^{\rm DPA}$ has an error of $\sim 10^{-3}\delta$. This is thought to be due to the damping-induced excitation and/or the fact that the DPA calculation was terminated up to $\mathcal D_2$. 
Although further refinements to incorporate these effects are available in principle, the overall features of the DC current are described well enough qualitatively and semi-quantitatively in our DPA.
In addition, it has a great advantage that the computational time required to obtain these results is almost the same as for the RTA.


{\it Conclusion}---
This letter highlights issues with the RTA for the DC current in insulating systems and proposes an improvement based on the QME. The RTA is frequently employed as a simple method, yet this phenomenological approach is exposed as a fatal approximation, displaying metallic DC conductivity in insulating systems within the linear response regime. 
This is because of
the incomplete inclusion of the first-order contribution of the electric field in the density matrix. 
We reevaluate the QME and incorporate the dynamical phase of the transition matrix, thereby correctly capturing the first-order terms of the electric field. We obtained a new scheme that accurately describes the insulating behavior. It was demonstrated that it accurately predicts the correct off-diagonal terms in the density matrix and is in semi-quantitative agreement with the numerically exact result. This also improves the calculation of quantum Hall effects. While numerically exact calculation can be performed for the present fundamental model, it is often time-consuming. Our method is not rigorous but correctly captures the overall features in the DC current. Furthermore, it does not require much computation time and is straightforward to apply to lattice systems. The DPA is an alternative to the RTA, which describes nonequilibrium steady states more correctly. We believe that this method will encourage progress in this research field. 

\begin{acknowledgments}
We are grateful to Y. Michishita, K. Takasan, A. Oguri, M. Sato, N. Kawakami, T. Morimoto and T. Oka for useful comments. This work was supported by KAKENHI Grants 
No.~19H01842, No.~19H05825.
SK is supported by JSPS KAKENHI Grant No. 20K14407.
\end{acknowledgments}


\clearpage
\appendix

\title{Supplemental Material: Unexpected linear conductivity in Landau-Zener model: limitations and improvements  of the relaxation time approximation in the quantum master equation}

\author{Ibuki Terada}
\affiliation{
Department of Physics, Ritsumeikan University, Shiga 525-8577, Japan
}
\author{Sota Kitamura}
\affiliation{
Department of Applied Physics, The University of Tokyo, Hongo, Tokyo, 113-8656, Japan
}
\author{Hiroshi Watanabe}
\affiliation{
Research Organization of Science and Technology, Ritsumeikan University, Shiga 525-8577, Japan
}
\author{Hiroaki Ikeda}
\affiliation{
Department of Physics, Ritsumeikan University, Shiga 525-8577, Japan
}

\date{\today}

\maketitle 

\renewcommand{\thesection}{\Alph{section}}
\renewcommand{\thefigure}{S\arabic{figure}}
\renewcommand{\theequation}{S\arabic{equation}}
\setcounter{figure}{0}
\setcounter{equation}{0}

\onecolumngrid

\section{Supplemental Material}
\section{Time evolution operator of isolated two-band systems}
We here introduce the snapshot basis~\cite{Kitamura,Kitamura2} as a solution of the time-dependent Schr\"odinger equation in the adiabatic limit (without the fermionic reservoir)~\cite{Davis,Dykhne}, in order to systematically expand the quantum master equation (QME) with respect to the external field (known as adiabatic perturbation theory).

According to the adiabatic theorem, the solution of the time-dependent Schr\"odinger equation 
in the adiabatic limit can be written as the instantaneous eigenstate of the system's Hamiltonian $H(k-Et)$ multiplied by a phase factor, 
\begin{equation}
\ket{\Phi_{k\alpha}(t)}=\ket{u_{k\alpha}(t)} e^{-i\Theta_{\alpha}(t)}.
\end{equation} 
Here we define the instantaneous eigenstate $\ket{u_{k\alpha}(t)}=\ket{u_{\alpha,k-Et}}$ with $H(k)|u_{\alpha,k}\rangle=\varepsilon_{k\alpha}|u_{\alpha,k}\rangle$, and the phase factor $\Theta_\alpha(t)$ given by
\begin{align}
\Theta_\alpha(t)&=\int_{t_0}^t dt'\,[\varepsilon_{k\alpha}(t')+EA_{\alpha\alpha}(t')]\\
&=\int_{t_0}^t dt'\,[\varepsilon_{k\alpha}(t')+E\mel{u_{k\alpha}(t')}{i\partial_k}{u_{k\alpha}(t')}],
\end{align}
which is introduced such that the diagonal element of the time-dependent Schr\"odinger equation vanishes.
The label $\alpha=\pm$ specifies the band index.

We expand the one-particle wave function $\ket{\Psi_k(t)}$ by the snapshot basis introduced above, as
\begin{align}\label{wave_func}
    \ket{\Psi_k(t)}=\sum_{\alpha=\pm}a_{k\alpha}(t)\ket{\Phi_{k\alpha}(t)}.
\end{align}
Then the time-dependent Schr\"odinger equation $i\partial_t \ket{\Psi_k(t)}=H(k-Et)\ket{\Psi_k(t)}$ can be rewritten as
\begin{align}\label{eq:EOM_a}
	i\frac{d}{dt}\left(\begin{matrix}a_{k+}(t)\\a_{k-}(t)\end{matrix}\right)
	&=\mathcal W_k(t)
	\left(\begin{matrix}a_{k+}(t)\\a_{k-}(t)\end{matrix}\right)\\
& =
\left(\begin{matrix}0&W_k(t)\\W_k^\ast(t)&0\end{matrix}\right)
\left(\begin{matrix}a_{k+}(t)\\a_{k-}(t)\end{matrix}\right)\\
& =
\left(\begin{matrix}0&E \mel{u_{k+}(t)}{i\partial_k}{u_{k-}(t)} e^{i\Theta_{+}(t)-i\Theta_{-}(t)}\\E \mel{u_{k-}(t)}{i\partial_k}{u_{k+}(t)} e^{i\Theta_{-}(t)-i\Theta_{+}(t)}&0\end{matrix}\right)
\left(\begin{matrix}a_{k+}(t)\\a_{k-}(t)\end{matrix}\right).
\end{align}

The formal solution of Eq.~\eqref{eq:EOM_a} is
$\left(a_{k+}(t), a_{k-}(t)\right)^T=U(t,t_0)\left(a_{k+}(t_0), a_{k-}(t_0)\right)^T$, where we introduce the time evolution operator (of the isolated system) in the snapshot basis,
\begin{equation}\label{eq:timeop}
    U(t,t_0)=\mathcal T\exp\left[-i\int_{t_0}^{t} \mathcal W_k(t')dt'\right],
\end{equation}
with the time-ordered product operator $\mathcal T$. 
One can show that $U(t,t_0)\to1$ for arbitrary $t$, when the system is gapped and the external field $E$ is infinitesimally small.

Here we also introduce the snapshot basis in the second-quantized form. 
We define the second-quantized Hamiltonian as
$\tilde H(k,t)=\sum_{\sigma\sigma'}\tilde c_{k\sigma}^\dag (t)[H(k-Et)]_{\sigma\sigma'}\tilde c_{k\sigma'}(t)$ with the field operator $\tilde c_{k\sigma}(t)$ in the Heisenberg representation.
Then the snapshot basis for the field operator is defined by~\cite{Kitamura2}
\begin{equation}
	\tilde\psi_{k\alpha}(t)
	=\sum_\sigma\braket{\Phi_{k\alpha}(t)}{\sigma}\tilde c_{k\sigma}(t).
\label{eq:unitra}
\end{equation}
The equation of motion for the snapshot field operator is obtained using $i\partial_t \tilde c_{k\sigma}(t)=[\tilde c_{k\sigma}(t),\tilde{H}(k,t)]$ as
\begin{equation}
	i\partial_t\tilde\psi_{k\alpha}(t)
	=\sum_{\beta}[\mathcal W_k(t)]_{\alpha\beta}\tilde\psi_{k\beta}(t),
\end{equation}
i.e., the first-quantized coefficients $a_{k\alpha}(t)$ and the second-quantized operators $\tilde\psi_{k\alpha}(t)$ obey the same equation of motion. It immediately follows that the snapshot field operator satisfies $\left(\tilde\psi_{k+}(t), \tilde\psi_{k-}(t)\right)^T=U(t,t_0)\left(\tilde\psi_{k+}(t_0), \tilde\psi_{k-}(t_0)\right)^T$.

\section{Density matrix within the RTA in the low $E$ regime }
Here we calculate the density matrix within the RTA at the low electric field. 
Diagonal and offdiagonal elements of Eq.~(\ref{eq:QME_origin}) with Eq.~(\ref{eq:coll_pheno}) read
\begin{align}
    \label{DE++}
    \frac{d[\rho_k(t)]_{\alpha\alpha}}{dt}&=-\alpha2\text{Im}\left[W^{\ast}_k(t)[\rho_k(t)]_{+-}\right]-\frac{[\rho_k(t)]_{\alpha\alpha}-f_D(\varepsilon_{k\alpha}(t))}{\tau_1},\\
    \label{DE+-}
    \frac{d[\rho_k(t)]_{+-}}{dt}&=-iW_k(t)\left[[\rho_k(t)]_{--}-[\rho_k(t)]_{++}\right]-\frac{[\rho_k(t)]_{+-}}{\tau_2}.
\end{align}
First (second) equation can be formally integrated when the offdiagonal (diagonal) matrix element is regarded as a source term. We obtain
\begin{align}
    \label{eq:sol++}
	[\rho_k(t)]_{\alpha\alpha}&=\int_{-\infty}^t\left\{ \frac{f_D(\varepsilon_{k\alpha}(s))}{\tau_1}-\alpha2\text{Im}\left[W^{\ast}_k(s)[\rho_k(s)]_{+-}\right]\right\} e^{-(t-s)/\tau_1}ds,\\
    \label{eq:sol+-}
	[\rho_k(t)]_{+-}&=-i\int_{-\infty}^{t}W_k(s)\left[[\rho_k(s)]_{--}-[\rho_k(s)]_{++}\right]e^{-(t-s)/\tau_2}ds,
\end{align}
where the dependence on the initial condition vanishes as we take the initial time as $t_0\rightarrow -\infty$. We can expand the first term in Eq.~\eqref{eq:sol++} as
\begin{equation}
\int_{-\infty}^t\frac{f_D(\varepsilon_{k\alpha}(s))}{\tau_1}e^{-(t-s)/\tau_1}ds=
f_D(\varepsilon_{k\alpha}(t))+E\tau_1\frac{\partial f_D(\varepsilon_{k\alpha}(t))}{\partial k}+\mathcal{O}(E^2),
\end{equation}
by performing the integral by parts. On the other hand, the second term in Eq.~\eqref{eq:sol++} has an $\mathcal O(E^2)$ contribution because 
$[\rho_k(s)]_{+-}=O(E)$ as we see below.
Therefore, the diagonal part of the density matrix is written as
\begin{equation}\label{eq:diagonal}
[\rho_k(t)]_{\alpha\alpha}\simeq f_D(\varepsilon_{k\alpha}(t))+E\tau_1\frac{\partial f_D(\varepsilon_{k\alpha}(t))}{\partial k}
\end{equation}
up to the first order of $E$, which corresponds to the Boltzmann transport theory. 
In a similar manner, we can expand Eq.~\eqref{eq:sol+-} with respect to $E$, using the relation
\begin{align}
e^{i\Theta_{+}(s)-i\Theta_{-}(s)+s/\tau_2}&=e^{i
\int_{t_0}^s dt'\,[\Delta_{k}(t')+E(A_{++}(t')-A_{--}(t'))]+s/\tau_2
}\\
&=e^{i\int_{t_0}^s dt'\,E(A_{++}(t')-A_{--}(t'))}
\frac{\frac{d}{ds}e^{i\int_{t_0}^s dt'\,\Delta_{k}(t')+s/\tau_2}}{i\Delta_{k}(s)+\tau_2^{-1}}
\end{align}
and performing the integral by parts. The result reads
\begin{equation}\label{eq:offdiagonal}
[\rho_{k}(t)]_{+-}\simeq-\dfrac{W_{k}(t)}{\Delta_{k}(t)-i\tau_{2}^{-1}}\left[f_{D}(\varepsilon_{k-}(t))-f_{D}(\varepsilon_{k+}(t))\right]
\end{equation}
at the first order of $E$ [Note that $W_k(t)=\mathcal{O}(E)$].

On the basis of above results \eqref{eq:diagonal} and \eqref{eq:offdiagonal}, we obtain the linear conductivity $\sigma^{\rm{RTA}}$,
\begin{align}
    \sigma^{\rm{RTA}}&=\sigma^{\rm{RTA}}_{\rm{intra}}
    +\sigma^{\rm{RTA}}_{\rm{inter}},\\
    \sigma^{\rm{RTA}}_{\rm{intra}}&=
    \tau_1\sum_{\alpha=\pm}\int\frac{dk}{2\pi}\left(\frac{\partial \varepsilon_{k\alpha}}{\partial k}\right)^2\left(-\frac{\partial f_D}{\partial \varepsilon_{k\alpha}}\right),\\
    \sigma^{\rm{RTA}}_{\rm{inter}}&=
   2\tau^{-1}_2\int\frac{dk}{2\pi}\frac{|\langle u_{+,k}|i\partial_k|u_{-,k}\rangle|^2\Delta_k}{\Delta_k^2+\tau_2^{-2}}\delta f_k.
\end{align}
In insulating systems, the intraband contribution $\sigma^{\rm{RTA}}_{\rm{intra}}$ vanishes because they have no Fermi surface. On the other hand, the interband contribution $\sigma^{\rm{RTA}}_{\rm{iner}}$ has finite value even in insulating systems and grows roughly in proportion to the interband damping $\tau_2^{-1}$.

The damping $i\tau_2$ in the denominator of Eq.~\eqref{eq:offdiagonal} also affects the Hall conductivity. 
The above formulation in 1D can be easily extended to 2D, with which the Hall conductivity can be calculated as
\begin{align}
    \frac{\sigma_{xy}-\sigma_{yx}}{2}&=\frac{e^2}{\hbar}\int\frac{d^2\bm{k}}{(2\pi)^2}\mathrm{Im}\left[\frac{\langle u_{+,\bm{k}}|i\partial_{k_x}|u_{-,\bm{k}}\rangle\langle u_{-,\bm{k}}|i\partial_{k_y}|u_{+,\bm{k}}\rangle\Delta_{\bm{k}}}{\Delta_{\bm{k}}-i\tau_2^{-1}}-\frac{\langle u_{+,\bm{k}}|i\partial_{k_y}|u_{-,\bm{k}}\rangle\langle u_{-,\bm{k}}|i\partial_{k_x}|u_{+,\bm{k}}\rangle\Delta_{\bm{k}}}{\Delta_{\bm{k}}-i\tau_2^{-1}}\right]\delta f_{\bm{k}}\\
    &=\frac{e^2}{\hbar}\int\frac{d^2\bm{k}}{(2\pi)^2}\frac{\Delta_{\bm{k}}^2}{\Delta_{\bm{k}}^2+\tau_2^{-2}}2\mathrm{Im}\left[
    \langle u_{+,\bm{k}}|i\partial_{k_x}|u_{-,\bm{k}}\rangle\langle u_{-,\bm{k}}|i\partial_{k_y}|u_{+,\bm{k}}\rangle
    \right]\delta f_{\bm{k}}\\
    &=-\frac{e^2}{h}\sum_{\alpha}\int\frac{d^2\bm{k}}{2\pi}\frac{\Delta_{\bm{k}}^2}{\Delta_{\bm{k}}^2+\tau_2^{-2}}\Omega_\alpha(\bm{k})f_{D}(\varepsilon_{\bm{k}\alpha}),
\end{align}
where we introduce the Berry curvature $\Omega_\alpha(\bm{k})=2\mathrm{Im}[\langle \partial_{k_x}u_{\alpha,\bm{k}}|\partial_{k_y}u_{\alpha,\bm{k}}\rangle]$. 

We recover the well-known TKNN formula for the anomalous Hall effect by taking $\tau_2^{-1}\to0$, which is known to be quantized in insulating systems at zero temperature.
While the nonzero damping $\tau_2^{-1}\neq0$ leads to the deviation from the quantized value in the RTA, this is also an artifact of the approximation as in the longitudinal conductivity derived above.

\section{The derivation of the quantum master equation}\label{derive_QME}
To reveal the dynamics of the two-band systems coupled to the fermionic reservoir, we here derive the QME in the snapshot basis microscopically. The QME determines the reduced density operator
$\hat \rho(t)=\mathrm{Tr_B}\hat \rho_{\rm{tot}}(t)$, where $\hat \rho_{\rm{tot}}(t)$ is the density matrix of the total system. If Hamiltonian is written as the sum of the momentum subspace $\hat H=\sum_k \hat H(k)$, we can construct the QME for the single electron with $k$.
We consider 
\begin{equation}\label{eq:QME_Appendix}
    \begin{split}
	   \frac{d\tilde \rho_{k}(t)}{dt}=-\int_{-\infty}^t
       \mathrm{Tr_B}\left[
	   \tilde H_{I,k}(t),\left[\tilde H_{I,k}(s),\tilde\rho_k(t)\otimes \tilde \rho_B(t_0)\right]\right]ds,
    \end{split}
\end{equation}
where $\tilde H_{I,k}$ is the interaction Hamiltonian between the system and the bath. Here the tilde symbol denotes the interaction picture based on the system decoupled from the reservoir. Equation~\eqref{eq:QME_Appendix} is known as the Redfield equation~\cite{Breuer}. 
We note that the following two approximations are employed in deriving the Redfield equation; (1) The Born approximation, i.e., the assumption that the coupling between the system and the bath is weak. (2) The Markov approximation, i.e., the assumption that the dynamics has no memory effect. 

For the fermionic reservoir employed in Ref.~\cite{Kitamura2}, the interaction Hamiltonian is given by
\begin{equation}\label{eq:Ham_int}
    \tilde H_{I,k}(t)
	=\sum_{\sigma,p}V_p\tilde b^{\dag}_{k\sigma p}(t)\tilde c_{k\sigma}(t)+\text {H.c.},
\end{equation}
where $\tilde c_{k\sigma}(t)$ is a fermion annihilation operator of an electron with $k$ and pseudospin $\sigma$, while $\tilde b_{k\sigma p}(t)=\tilde b_{k\sigma p}(0)e^{-i\omega_p t}$ is an annihilation operator of an electron in the fermionic reservoir whose mode energy is $\omega_p$. 

Let us compute the partial trace in Eq.~\eqref{eq:QME_Appendix} using the explicit expression Eq.~\eqref{eq:Ham_int}. Before that, we introduce the snapshot basis for the field operator $\tilde{\psi}_{k\alpha}(t)$ defined as Eq.~\eqref{eq:unitra},
in order to express the $s$ integral in a form convenient for the adiabatic perturbation theory.
We can rewrite the QME as
\begin{equation}\label{eq:QMM0}
	\begin{split}
		\frac{d\tilde \rho_{k}(t)}{dt}&=
		\sum_{\alpha,\beta}\int_{0}^\infty ds\ \mathcal{C}_{\rm p}(s)
		\braket{\Phi_{k\beta}(t-s)}{\Phi_{k\alpha}(t)}\left[\tilde \psi^{\dag}_{k\beta}(t-s)\tilde \rho_{k}(t),\tilde \psi_{k\alpha}(t)\right]+\text{H.c.}\\
		&+\sum_{\alpha,\beta}\int_{0}^\infty ds\ \mathcal{C}_{\rm h}(s)
		\braket{\Phi_{k\alpha}(t)}{\Phi_{k\beta}(t-s)}\left[\tilde \psi_{k\beta}(t-s)\tilde \rho_{k}(t),\tilde \psi^{\dag}_{k\alpha}(t)\right]+\text{H.c.},
	\end{split}
\end{equation}
where we introduced the particle/hole time correlation function of the fermionic reservoir~\cite{DelRe,Kitamura2}
\begin{align}
\mathcal{C}_{\rm p}(s) & =\sum_{p}|V_{p}|^{2}\text{Tr}_{\text{B}}\left[\tilde{b}_{k\sigma p}^{\dagger}(t)\tilde{b}_{k\sigma p}(t-s)\tilde{\rho}_{B}(t_{0})\right],\\
\mathcal{C}_{\rm h}(s) & =\sum_{p}|V_{p}|^{2}\text{Tr}_{\text{B}}\left[\tilde{b}_{k\sigma p}(t)\tilde{b}_{k\sigma p}^{\dagger}(t-s)\tilde{\rho}_{B}(t_{0})\right].
\end{align}
Using the broadband condition Eq.~\eqref{eq:BBC}, we can explicitly calculate the correlation function as
\begin{align}
\mathcal{C}_{\rm h}(s)=\mathcal{C}_{\rm p}(s) & =2\Gamma\int\frac{d\omega}{2\pi}e^{i\omega s}f_{D}(\omega)
  =\Gamma\left(\delta(s)-i\frac{T}{\sinh(\pi Ts)}\right),\label{eq:time_corre}
\end{align}
which decays with a decay time $\tau_B=1/\pi T$ ($k_B=1$)~\cite{DelRe}. Substituting Eq.~\eqref{eq:time_corre} into Eq.~\eqref{eq:QMM0}, we obtain
\begin{equation}\label{eq:QME1}
	\begin{split}
		\frac{d\tilde \rho_{k}(t)}{dt}&=
		\Gamma\sum_{\alpha}\left(\tilde \psi^{\dag}_{k\alpha}(t)\tilde \rho_{k}(t)\tilde \psi_{k\alpha}(t)+\tilde \psi_{k\alpha}(t)\tilde \rho_{k}(t)\tilde \psi^{\dag}_{k\alpha}(t)-\tilde \rho_{k}(t)\right)\\
		&-\Gamma\sum_{\alpha,\beta}\int_{0}^\infty ds\ \frac{iT}{\sinh(\pi Ts)}
		\braket{\Phi_{k\alpha}(t)}{\Phi_{k\beta}(t-s)}\left[\left\{\tilde \rho_{k}(t),\tilde \psi_{k\beta}(t-s)\right\},\tilde \psi^{\dag}_{k\alpha}(t)\right]+\text{H.c.}
	\end{split}
\end{equation}
Note that we used $\int_0^\infty ds\delta(s)=1/2$ for the first term.

Due to the decaying nature of the correlation function $\mathcal{C}_{\rm p/h}(s)$, the integrand of Eq.~\eqref{eq:QME1} only need to be accurate in $s \lesssim \tau_B$. Here we assume that the change in the momentum $\Delta k=Es$ is small enough in $s \lesssim \tau_B$, with which one can approximate $\ket{\Phi_{k\alpha}(t-s)}$ and $\tilde\psi_{k\alpha}(t-s)$ using the Taylor expansion.

Let us start with the expansion of the phase factor $e^{-i\Theta_{\alpha}(t)}=e^{-i\int_{t_0}^{t}dt'\left[\varepsilon_{k\alpha}(t')+EA_{\alpha\alpha}(t')\right]dt'}$.
The exponent at $t-s$ is expanded as
\begin{equation}
    -i\int_{t_0}^{t-s}dt'\left[\varepsilon_{k\alpha}(t')+EA_{\alpha\alpha}(t')\right]dt'=-i\int_{t_0}^tdt'\left[\varepsilon_{k\alpha}(t')+EA_{\alpha\alpha}(t')\right]dt'+i\varepsilon_{k\alpha}(t)s+iEsA_{\alpha\alpha}(t)
    +\frac{iEs^2}{2}\frac{\partial \varepsilon_{k\alpha}(t)}{\partial k}+\cdots.
\end{equation}
The third and subsequent terms vanish in the adiabatic limit, 
so that here we only take account of them in the leading order in $\Delta k=Es$. Then the phase factor is approximated as
\begin{equation}\label{eq:DPA}
    e^{-i\Theta_{\alpha}(t-s)}
    \simeq\left(1+iEsA_{\alpha\alpha}(t)\right)e^{-i\Theta_{\alpha}(t)+i\varepsilon_{k\alpha}(t)s}
\end{equation}
where we use $e^{iEsA_{\alpha\alpha}(t)}\simeq 1+iEsA_{\alpha\alpha}(t)$. Because the factor $e^{+i\varepsilon_{k\alpha}(t)s}$ plays an important role in the later discussion, we here call this approximation the dynamical phase approximation (DPA). 

In a similar manner, we can expand the instantaneous eigenstate $\ket{u_{k\alpha}(t-s)}$ as 
\begin{align}
    \ket{u_{k\alpha}(t-s)}&\simeq \ket{u_{k\alpha}(t)}-s\frac{d}{dt}\ket{u_{k\alpha}(t)}\\
    &=\ket{u_{k\alpha}(t)}-iEs\sum_{\beta}A_{\beta\alpha}(t)\ket{u_{k\beta}(t)},
\end{align}
where the terms with the second or higher order of $s$ were dropped. 
With these, we can evaluate the snapshot basis $\ket{\Phi_{k\alpha}(t)}=\ket{u_{k\alpha}(t)}e^{-i\Theta_\alpha(t)}$ at $t-s$ with the DPA as 
\begin{equation}\label{snap_DPA}
\begin{split}
    \ket{\Phi_{k\alpha}(t-s)}&\simeq \left(1+iEsA_{\alpha\alpha}\right)e^{-i\Theta_{\alpha}(t)+i\varepsilon_{k\alpha}(t)s}\left(\ket{u_{k\alpha}(t)}-iEs\sum_{\beta}A_{\beta\alpha}(t)\ket{u_{k\beta}(t)}\right)\\
    &\simeq e^{+i\varepsilon_{k\alpha}(t)s}\left(\ket{\Phi_{k\alpha}(t)}-is\sum_{\beta}\left[\mathcal W_k(t)\right]_{\beta\alpha}\ket{\Phi_{k\beta}(t)}\right).
\end{split}
\end{equation}

Next, let us consider the snapshot basis operator $\tilde\psi_{k\alpha}(t-s)$ with the DPA. Since the snapshot basis operator evolves as $\tilde{\psi}_{k\alpha}(t)=\sum_\beta[U(t,t')]_{\alpha\beta}\tilde{\psi}_{k\beta}(t')$ with Eq.~\eqref{eq:timeop}~\cite{Kitamura2}, the quantity we need to approximate here is 
\begin{equation}\label{eq:U_expnsion}
    \left[U(t-s,t)\right]_{\alpha\beta}=\delta_{\alpha\beta}-i\int_{0}^{-s}dt_1\left[\mathcal W_k(t+t_1)\right]_{\alpha\beta}
    +(-i)^2\int_{0}^{-s}dt_1\int_{0}^{t_1}dt_2\sum_{\gamma}\left[\mathcal W_k(t+t_1)\right]_{\alpha\gamma}\left[\mathcal W_k(t+t_2)\right]_{\gamma\beta}+\cdots.
\end{equation}
In order to reproduce the first order of $Es$, it is sufficient to truncate at the second term and approximate the integrand as
\begin{equation}\label{eq:W_app}
    \left[\mathcal W_k(t+t_1)\right]_{\alpha\beta}\simeq\left[\mathcal W_k(t)\right]_{\alpha\beta}e^{\alpha i\Delta_k(t)t_1},
\end{equation}
according to Eq.~\eqref{eq:DPA}. However, this approximation turns out to show unphysical behavior due to the broken unitarity of $U(t,t')$ upon the truncation. Also, we note that the norm of $\ket{\Phi_{k\alpha}(t-s)}$ is not conserved in this treatment.

To remedy this problem, we here include the third term in Eq.~\eqref{eq:U_expnsion} while retaining the use of Eq.~\eqref{eq:W_app}, which results in
\begin{equation}\label{eq:U_app}
    \left[U(t-s,t)\right]_{\alpha\beta}\simeq\delta_{\alpha\beta}-\alpha\left[\mathcal W_k(t)\right]_{\alpha\beta}\frac{e^{-\alpha i\Delta_k(t)s}-1}{\Delta_k(t)}
    +\frac{|W_k(t)|^2}{\Delta_k(t)}\left(i\alpha s+\frac{e^{-\alpha i \Delta_k(t)s}-1}{\Delta_k(t)}\right)\delta_{\alpha\beta}.
\end{equation}
With this approximation, the transformed operator 
\begin{align}
\tilde\psi_{k\alpha}'(t,s)&=\sum_{\beta}\braket{\Phi_{k\alpha}(t)}{\Phi_{k\beta}(t-s)}\tilde\psi_{k\beta}(t-s)
\end{align}
appearing in Eq.~\eqref{eq:QME1} is approximated as 
\begin{align}
\tilde\psi_{k\alpha}'(t,s)&\simeq\sum_\beta e^{i\varepsilon_{k\alpha}(t)s}\left[\delta_{\alpha\beta}+\left(-
\alpha\left[\mathcal W_k(t)\right]_{\alpha\beta}+ \frac{|W_k(t)|^2}{\Delta_k(t)}\delta_{\alpha\beta}\right)
\left(i\alpha se^{-\alpha i\Delta_{k}(t)s}+\frac{e^{-\alpha i\Delta_{k}(t)s}-1}{\Delta_k(t)}\right)
\right]\tilde\psi_{k\beta}(t),\label{eq:DPA-operator}
\end{align}
which satisfies
\begin{align}
\left\{\tilde\psi_{k\alpha}'(t,s),\tilde\psi_{k\beta}'^\dag(t,s)\right\}&=
\left(1+|W_k(t)|^2s^2\right)\delta_{\alpha\beta}+\mathcal{O}(E^3).
\end{align}
This implies that the unitary nature of the time evolution of the isolated system is maintained up to the first order of $s$ and the second order of $E$.

With the use of Eq.~\eqref{eq:DPA-operator} and
\begin{align}
    \int_0^\infty \frac{T\sin(\varepsilon s)}{\sinh(\pi Ts)}ds
    &=-f_D(\varepsilon)+\frac{1}{2},\\    
    -i\int_0^\infty \frac{Ts^le^{-i\varepsilon s}}{\sinh(\pi Ts)}ds
    &=i^l\frac{\partial^l}{\partial \varepsilon^l}\left[f_D(\varepsilon)
    +\frac{i}{\pi}\mathrm{Re}\Psi\left(\frac{1}{2}+\frac{i\varepsilon}{2\pi T}\right)\right]\quad(l\ge1),
\end{align}
we can perform the integral in Eq.~\eqref{eq:QME1}.
Here $\Psi$ is the digamma function. 
The resultant equation is given by
\begin{equation}\label{eq:QME2}
\begin{split}
\frac{d\tilde \rho_{k}(t)}{dt}&\simeq \tilde{\mathcal{D}}_{0}(t)+\tilde{\mathcal{D}}_{1}(t)+\tilde{\mathcal{D}}_{2}(t),\\
\tilde{\mathcal{D}}_{0}(t)&=
2\Gamma\sum_{\alpha}f_D(\varepsilon_{k\alpha}(t))\left(
\tilde \psi^{\dag}_{k\alpha}(t)\tilde \rho_{k}(t)\tilde\psi_{k\alpha}(t)-\frac{1}{2}\left\{\tilde \rho_{k}(t),\tilde\psi_{k\alpha}(t)\tilde \psi^{\dag}_{k\alpha}(t)\right\}\right)\\
&+2\Gamma\sum_{\alpha}(1-f_D(\varepsilon_{k\alpha}(t)))\left(\tilde \psi_{k\alpha}(t)\tilde \rho_{k}(t)\tilde \psi^{\dag}_{k\alpha}(t)-\frac{1}{2}\left\{\tilde \rho_{k}(t),\tilde \psi^{\dag}_{k\alpha}(t)\tilde\psi_{k\alpha}(t)\right\}\right),\\
\tilde{\mathcal{D}}_{1}(t)&=
2\Gamma\sum_{\alpha\beta}\frac{\left[\mathcal W_k(t)\right]_{\alpha\beta}}{\Delta_k(t)}
\left(\delta f_k(t)+\frac{\Delta_k(t)}{2}
\frac{\partial}{\partial \varepsilon}\left[f_D(\varepsilon_{k,-}(t))
+f_D(\varepsilon_{k,+}(t))
-\frac{i\alpha}{\pi}\delta\Psi_k(t)
\right]
\right)
\\
&\times\left(
\tilde\psi_{k\beta}(t)\tilde \rho_{k}(t)\tilde \psi^{\dag}_{k\alpha}(t)
-\tilde \psi^{\dag}_{k\alpha}(t)\tilde \rho_{k}(t)\tilde\psi_{k\beta}(t)
-\left\{\tilde \rho_{k}(t),\tilde \psi^{\dag}_{k\alpha}(t)\tilde\psi_{k\beta}(t)\right\}
\right),\\
\tilde{\mathcal{D}}_{2}(t)&=
-2\Gamma\sum_{\alpha}\alpha\frac{|W_k(t)|^2}{\Delta_k^2(t)}\left(
\delta f_k(t)
+\Delta_k(t)
\frac{\partial f_D(\varepsilon_{k,-\alpha}(t))}{\partial \varepsilon}
\right)
\\
&\times
\left(
\tilde \rho_{k}(t)
+\tilde\psi_{k\alpha}(t)\tilde \rho_{k}(t)\tilde \psi^{\dag}_{k\alpha}(t)
-\tilde \psi^{\dag}_{k\alpha}(t)\tilde \rho_{k}(t)\tilde\psi_{k\alpha}(t)
-\left\{\tilde \rho_{k}(t),\tilde \psi^{\dag}_{k\alpha}(t)\tilde\psi_{k\alpha}(t)\right\}
\right),
\end{split}
\end{equation}
where $\delta \Psi_k(t)=\mathrm {Re}\Psi(\frac{1}{2}-\frac{i\beta\varepsilon_{k-}(t)}{2\pi})-\mathrm {Re}\Psi(\frac{1}{2}-\frac{i\beta\varepsilon_{k+}(t)}{2\pi})$.

We redefine the density matrix in the snapshot basis as $[\rho_k(t)]_{\alpha\beta}=\mathrm{Tr}[\tilde \psi^{\dag}_{k\beta}(t)\tilde \psi_{k\alpha}(t)\tilde\rho_k(t)]$. We obtain the equation of motion for the matrix element of $\rho_k(t)$ as
\begin{equation}\label{eq:EOM_p}
    \frac{d[\rho_k(t)]_{\alpha\beta}}{dt}=-i\left[[\mathcal W_k(t),\rho_k(t)]\right]_{\alpha\beta}+\mathrm{Tr}\left[\tilde \psi^{\dag}_{k\beta}(t)\tilde \psi_{k\alpha}(t)\frac{d\tilde\rho_k(t)}{dt}\right].
\end{equation}
The second term of the right hand side in Eq.~\eqref{eq:EOM_p} corresponds to the dissipation term. Substituting Eq.~\eqref{eq:QME2} into Eq.~\eqref{eq:EOM_p}, we finally arrive at
\begin{align}
[{\mathcal{D}}_{0}(t)]_{\alpha\beta} & =-2\Gamma([\rho_{k}(t)]_{\alpha\beta}-f_{D}(\varepsilon_{k\alpha}(t))\delta_{\alpha\beta}),\\{}
[{\mathcal{D}}_{1}(t)]_{\alpha\beta} & =-2\Gamma\frac{\left[\mathcal{W}_{k}(t)\right]_{\alpha\beta}}{\Delta_{k}(t)}\left(\delta f_{k}(t)+\frac{\Delta_{k}(t)}{2}\frac{\partial}{\partial\varepsilon}\left[f_{D}(\varepsilon_{k,-}(t))+f_{D}(\varepsilon_{k,+}(t))-\frac{i\alpha}{\pi}\delta\Psi_{k}(t)\right]\right),\\{}
[{\mathcal{D}}_{2}(t)]_{\alpha\beta} & =2\Gamma\alpha\frac{|W_{k}(t)|^{2}}{\Delta_{k}^{2}(t)}\left(\delta f_{k}(t)+\Delta_{k}(t)\frac{\partial f_{D}(\varepsilon_{k,-\alpha}(t))}{\partial\varepsilon}\right)\delta_{\alpha\beta}
\end{align}
where $\mathcal{D}_{n}$ is the $n$-th order of $E$ term in the dissipation term. In the presence of the particle-hole symmetry, the derivative term in $\mathcal{D}_{1}$ vanishes since $f_D(\varepsilon_{k+}(t))+f_D(\varepsilon_{k-}(t))=1$ and $\delta\Psi_k(t)=0$.

\end{document}